\documentclass[referee,10pt]{aa}
\usepackage{graphicx}
\usepackage{lscape}
\usepackage{epsfig}
\begin{document}

\title{Discovery of eleven new ZZ Ceti stars{\footnote{Partially based on 
observations at Observat\'orio do Pico dos Dias/LNA, the Southern Astrophysical 
Research telescope, a collaboration between CNPq-Brazil, NOAO, UNC and MSU, and
McDonald Observatory of the University of Texas at Austin.}}
}

\author{
B. G. Castanheira\inst{1}\inst{,2},
S. O. Kepler\inst{1},
F. Mullally\inst{2},
D. E. Winget\inst{2},
D. Koester\inst{3},
B. Voss\inst{3},
S. J. Kleinman\inst{4},
A. Nitta\inst{4},
D. J. Eisenstein\inst{5},
R. Napiwotzki\inst{6},
D. Reimers\inst{7}}

\offprints{barbara@if.ufrgs.br}
\institute{
Instituto de F\'{\i}sica, Universidade Federal do Rio Grande do Sul,
  91501-900  Porto-Alegre, RS, Brazil\\
\and Department of Astronomy and McDonald Observatory,
  University of Texas,
  Austin, TX 78712, USA\\
\and Institut f\"ur Theoretische Physik und Astrophysik, Universit\"at Kiel,
  24098 Kiel, Germany
\and Sloan Digital Sky Survey, Apache Pt. Observatory, PO Box 59, Sunspot, NM
  88349, USA
\and Steward Observatory, University of Arizona\\
933 N. Cherry Ave.  Tucson, AZ 85721, USA\\
\and Centre for Astrophysics Research, University of
  Hertfordshire, College Lane, Hatfield AL10 9AB, UK
\and Hamburger Sternwarte, Gojensbergweg 112, D-21029 Hamburg,
  Germany
\\}

\date{Received --; accepted --}

\abstract{We report the discovery of eleven new ZZ Cetis using telescopes
at OPD (Observat\'orio do Pico dos Dias/LNA) in Brazil, the 4.1~m SOAR
(Southern Astrophysical Research) telescope at Cerro Pachon, Chile, and the
2.1~m Otto Struve telescope at McDonald observatory. The candidates were
selected from the SDSS (Sloan Digital Sky Survey) and SPY (ESO SN Ia progenitor
survey), based on their $T_{\mathrm{eff}}$ obtained from optical spectra  
fitting. This selection criterion yields the highest success rate of 
detecting new ZZ Cetis, above 90\% in the $T_{\mathrm{eff}}$ range from 
12\,000 to 11\,000\,K. We also report on a DA not observed to vary, with a
$T_{\mathrm{eff}}$ placing the star close to the blue edge of the instability 
strip. Among our new pulsators, one is a little bit cooler than this star
for which pulsations were not detected. Our observations are an important 
constraint on the 
location of the blue edge of the ZZ Ceti instability strip.
}

\titlerunning{Eleven new ZZ Cetis}
\authorrunning{B. G. Castanheira  et al.}

\maketitle


\section{Introduction}

White dwarf stars are the end point of the evolution of 95-98\% of all stars.
As they cool, white dwarf stars pass through three distinct instability strips, 
depending on their temperatures, atmospheric composition, and the element that
drives pulsation: carbon and/or oxygen in DOVs, helium in DBVs, or hydrogen 
in DAVs 
or ZZ Cetis. Observationally, the DA instability strip ranges in effective 
temperature ($T_{\mathrm{eff}}$) from 12\,270 to 10\,850\,K (Bergeron et al.
2004, Mukadam et al. 2004, and Gianninas et al. 2005). While Bergeron's 
instability strip is pure,
Mukadam et al. (2004) and Mullally et al. (2005) did not detect
light variability in a significant number of stars located within the 
$T_{\mathrm{eff}}$ and $\log g$ boundaries of the instability strip.
If the instability strip is contaminated (occupied by non-pulsators of the 
same spectral type as pulsators), pulsation would no longer be a normal stage
of white dwarf 
evolution. However, because of destructive interference between modes, some
white dwarf stars can appear to be constant for timescales of 3-4 hours.
Another important issue is that some known pulsating white
dwarf stars show amplitudes as low as 4 mma (e.g. Kanaan et al.
2005), so we have to reach {\it at least} this limit to determine
whether a star pulsates or not; many stars not
observed to vary by Mukadam et al. 2004 have detection limits above 4 mma.

Pulsations are global; each periodicity is an independent measurement 
of the interior. This gives rise to asteroseismology: the study of stars below
the photosphere through measurement of their pulsations.
Comparing the observed pulsation spectrum to models of the stellar interior
through asteroseismology allows for the stellar mass determination 
(Winget et al. 1990, 
Bradley \& Winget 1994) and even constrain the value of the
C$^{12}$($\alpha$,$\gamma$)O$^{16}$ cross section (e.g. Metcalfe et al. 2002),
which can only be measured in a terrestrial 
laboratory at energy levels eight orders of magnitude higher. Furthermore,
the cooling time
scales of DAVs (Kepler et al. 2000, 2005, Mukadam et al. 2003) can be used to 
calibrate the age of the galactic morphological components (Winget et al. 1987,
Hansen et al. 2002, von Hippel 2005), by observing field stars, open and 
globular clusters up
to magnitudes to include the turnoff of the white dwarf cooling sequence.

Pulsating white dwarf stars are also important for the study of extreme
physics: internal crystallization (Winget et al. 1997, Kanaan et al. 2005), 
neutrino cooling (Kawaler et al. 1986, Winget et al. 2004) and axion emission  
(C\'orsico et al. 2001, Kepler 2004, Kim et al. 2005). Both particles can be 
created at temperatures and densities found in white dwarf cores and 
consequently must be taken into account in the cooling models.

Quasar surveys, like SDSS (Sloan Digital Sky Survey), HE (Hamburg
ESO) and 2dF (Two Degree Field), are increasing the number of spectroscopically
identified white dwarf stars as a by-product; quasars and white dwarf stars
have similar colors. As these new white dwarf stars are fainter than the 
ones previously known, but with similar $T_{\mathrm{eff}}$, we are 
able to study a sample of stars at larger distances, with a wider range of
progenitor metallicity. The study of their chemical composition
is of interest if we are to use SN Ia as standard candles, that are 
products of mass accretion onto white dwarf stars.

Mukadam et al. (2004) show that at least 90\% of the SDSS candidates with
12\,000$\geq T_{\mathrm{eff}} \geq$11\,000\,K are pulsators. 
We selected our targets based on the $T_{\mathrm{eff}}$ we derive by
fitting the whole optical spectra, as described by Kleinman et al. (2004).
For the candidates with spectra obtained by the SPY survey  (e.g. Napiwotzki et
al. 2003), we derive $T_{\mathrm{eff}}$ by fitting only the hydrogen 
line profiles. 
In both fitting procedures, we use the same model atmosphere grid,
developed by Detlev Koester (similar to that described in 
Finley et al. 1997).
We only obtained time series photometry of stars with $T_{\mathrm{eff}}$
inside or close to the ZZ Ceti instability strip. All observed stars, but one,
turned out to be pulsators.

In this paper, we report the discovery of eleven new pulsators and one star not
observed to vary down to a detection limit of 3.75\,mma.

\section{Observations}

We observed some of our candidates at Observat\'orio Pico dos Dias, LNA, in 
Brazil. For these observations, we used the 1.6-m telescope, with a frame 
transfer CCD 301, focal reducer, and no filters. The CCD has a quantum 
efficiency of 60\% at around 4000\AA, the prefered wavelength range to observe
ZZ Cetis to maximize the pulsation amplitude versus CCD efficiency. The 
integration times range from 15 to 45 seconds, depending on readout times,
weather conditions, and apparent stellar magnitude. We observed without filters
because the g--mode pulsations in ZZ Ceti stars are coherent at all optical
wavelengths (Robinson, Kepler \& Nather 1982). We also observed bright 
candidates with the 0.6-m Zeiss telescope. We used the CCD 106, which has a 
7.5~s readout,
and integrations of 30 to 60 seconds.

In June and July 2005, we used the SOAR Optical Imager, a mosaic of two EEV
2048$\times$4096 CCDs, thinned and back illuminated,
with an efficiency around 73\% at 4000\,\AA , at the cassegrain focus
of the 4.1-m SOAR telescope. We observed in fast readout mode, with 
the CCDs binned 4$\times$4, to decrease the
readout+write time to 6.4\,s, and still achieve a 0.354"/pixel resolution. 
The exposure times ranged from 20 to 40\,s.
All observations were obtained with a Johnson B filter.

We also observed with the Otto Struve 2.1-m telescope at McDonald Observatory,
using the Argos camera, a frame transfer CCD in the prime focus, 
with a BG40 filter to reduce the scatter from sky (Nather \& Mukadam 2004).
The integration times were 20 and 30\,s.

In Table~\ref{log1}, we show the journal of observations for all
stars. 

\begin{table}[h]
\begin{tabular}{||c|c|c|c|c|c||}\hline\hline
Star & Run start (UT) & $t_{\mathrm{exp}}$ (s) & $\Delta T$ (hr) & \# points &
Telescope\\ 
\hline
HE~0031-5525 & 2004-08-14 04:03 & 40 & 0.71 & 6464  & 1.6m\\
 & 2004-08-18 05:09 & 40 & 3.53 & 253 & 0.6m\\
 & 2004-08-19 04:00 & 30 & 4.80 & 418 & 0.6m\\
 & 2004-09-10 03:59 & 20 & 2.34 & 422 & 1.6m\\
 & 2004-09-11 02:31 & 30 & 0.88 & 105 & 1.6m\\
SDSS J024922.3-010006.7 & 2004-09-11 04:35 & 40 & 3.66 & 329 & 1.6m\\
 & 2004-11-13 01:56 & 20 & 2.44 & 439 & 1.6m\\
 & 2004-11-14 03:16 & 30 & 3.52 & 422 & 1.6m\\
 & 2004-11-15 00:30 & 15 & 2.56 & 615 & 1.6m\\
SDSS J125710.5+012422.9 & 2005-06-05 00:32 & 20 & 1.70 & 234 & 4.1m\\
 & 2005-06-07 00:03 & 30 & 2.00 & 193 & 4.1m\\
SDSS J153332.9-020600.0 & 2004-08-10 22:06 & 20 & 1.95 & 351 & 1.6m\\
 & 2004-08-11 21:56 & 20 & 1.81 & 325 & 1.6m\\
SDSS J161837.2-002302.7 & 2005-06-06 02:59 & 40 & 4.13 & 321 & 4.1m\\
 & 2005-06-07 05:54 & 40 & 2.00 & 160 & 4.1m \\
SDSS J164115.5+352140.6 & 2005-06-06 08:15 & 20 & 2.42 & 436 & 2.1m \\
 & 2005-06-09 07:14 & 30 & 0.40 & 48 & 2.1m \\
 & 2005-06-09 09:35 & 30 & 1.15 & 138 & 2.1m \\
SDSS J212808.4-000750.8 & 2004-08-11 00:20 & 30 & 2.15 & 258 & 1.6m\\
 & 2004-08-12 01:33 & 30 & 0.74 & 89 & 1.6m\\
SDSS J213530.3-074330.7 & 2004-08-11 02:43 & 30 & 1.71 & 205 & 1.6m\\
 & 2004-08-12 02:20 & 30 & 1.06 & 127 & 1.6m\\
SDSS J214723.7-001358.4 & 2005-07-05 08:14 & 40 & 2.10 & 115 & 4.1m\\
 & 2005-07-07 05:25 & 30 & 3.00 & 300 & 4.1m\\
SDSS J215354.1-073121.9 & 2005-06-06 07:33 & 30 & 2.00 & 200 & 4.1m\\
 & 2005-06-07 08:36 & 30 & 2.00 & 200 & 4.1m\\ 
SDSS J223135.7+134652.8 & 2004-08-13 01:34 & 40 & 3.88 & 349 & 1.6m\\
 & 2004-08-14 01:48 & 40 & 1.93 & 174 & 1.6m\\
 & 2004-08-18 01:26 & 60 & 1.58 & 82 & 0.6m\\
 & 2004-08-19 01:54 & 45 & 1.65 & 107& 0.6m\\
 & 2004-09-10 00:30 & 30 & 3.14 & 377 & 1.6m\\
 & 2004-09-10 23:57 & 30 & 2.43 & 292 & 1.6m\\
SDSS J230726.6-084700.2 & 2004-08-11 04:32 & 40 & 1.28 & 115 & 1.6m\\
\hline\hline
\end{tabular}
\caption{Journal of observations with the 1.6-m and 0.6-m telescope at OPD,
the 4.1-m SOAR telescope, and the 2.1-m telescope at McDonald observatory.
$\Delta T$ is the length of the observing run and 
$t_{\mathrm{exp}}$ is the integration time of each exposure.}
\label{log1}
\end{table}

The technique to detect variability is differential time series photometry, 
comparing the targets with the other stars in the same field, to
minimize the effects of sky and transparency fluctuations.

For each run, we extracted light curves using the IRAF script hsp, developed by 
Antonio Kanaan, using different aperture sizes and select the light curve with
either the lowest noise level or the highest signal--to--noise rate (SNR)
in the Fourier transform. 

\section{Results}

In Table~\ref{newv}, we list the new variable stars and their physical 
parameters
derived from optical spectra, using an improved model grid similar
to that described in
Finley at al. (1997). We fit the candidates from SDSS using the whole
optical spectra and photometry, as in Kleinman et al. (2004), while for the 
candidate from HE we used only
the hydrogen lines in the optical spectra, because it is independent of the
flux calibration and the SNR is high enough.

\begin{table}
\begin{tabular} {|c|c|c|c|c|c|}
\hline
Star & RA (2000) & DEC (2000) & $T_{\mathrm{eff}}$ (K) & $\log g$ & g (mag) 
\\\hline \hline
HE 0031-5525 & 00:33:36 & -55:08:37 & 11480$\pm$30 & 7.65$\pm$0.02 & 15.94 \\
SDSS J024922.3-010006.7 & 02:49:22.3 & -01:00:06.7 & 11060$\pm$110 & 
8.31$\pm$0.10 & 19.08 \\
SDSS J125710.5+012422.9 & 12:57:10.5 & +01:24:22.9 & 11520$\pm$160 &
8.36$\pm$0.09 & 18.65 \\
SDSS J153332.9-020600.0 & 15:33:32.9 & -02:06:00.0 & 11350$\pm$40 & 
8.20$\pm$0.02 & 16.62 \\
SDSS J161837.2-002302.7 & 16:18:37.2 & -00:23:02.7 & 10860$\pm$160 &
8.16$\pm$0.12 & 19.26 \\
SDSS J164115.5+352140.6 & 16:41:15.5 & +35:21:40.6 & 11230$\pm$160 &
8.43$\pm$0.10 & 19.04 \\
SDSS J212808.4-000750.8 & 21:28:08.4 & -00:07:50.8 & 11440$\pm$100 & 
8.29$\pm$0.07 & 17.97 \\
SDSS J213530.3-074330.7 & 21:35:30.3 & -07:43:30.7 & 11190$\pm$120 & 
7.67$\pm$0.09 & 18.59 \\
SDSS J215354.1-073121.9 & 21:53:54.1 & -07:31:21.9 & 11930$\pm$130 &
8.07$\pm$0.06 & 18.45 \\
SDSS J223135.7+134652.8 & 22:31:35.7 & +13:46:52.8 & 11080$\pm$100 & 
7.95$\pm$0.07 & 18.63 \\
SDSS J230726.6-084700.2 & 23:07:26.6 & -08:47:00.2 & 11060$\pm$110 & 
8.19$\pm$0.09 & 18.83\\
\hline
\end{tabular}
\caption{List of properties of
the new variables. The $T_{\mathrm{eff}}$ and $\log g$ were
derived from optical spectra, with the same model grid, despite different
fitting procedures for HE and SDSS stars.}
\label{newv}
\end{table}

\begin{figure}
\centering
\includegraphics[angle=0,width=\linewidth]{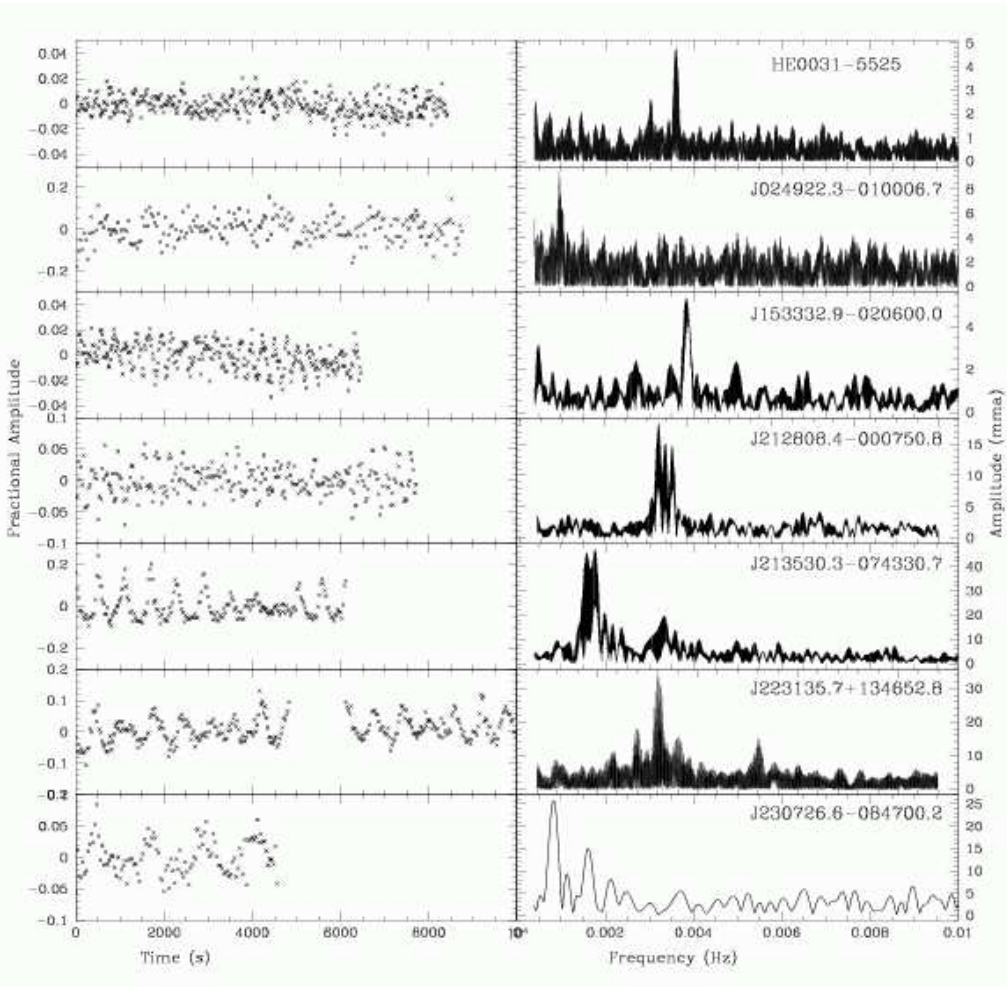}
\caption{The light curves of a single run (left panels) and the Fourier 
transforms of the complete data set
(right panels) for the new ZZ Cetis, discovered at LNA.} 
\label{lc}
\end{figure}

\begin{figure}
\centering
\includegraphics[angle=0,width=\linewidth]{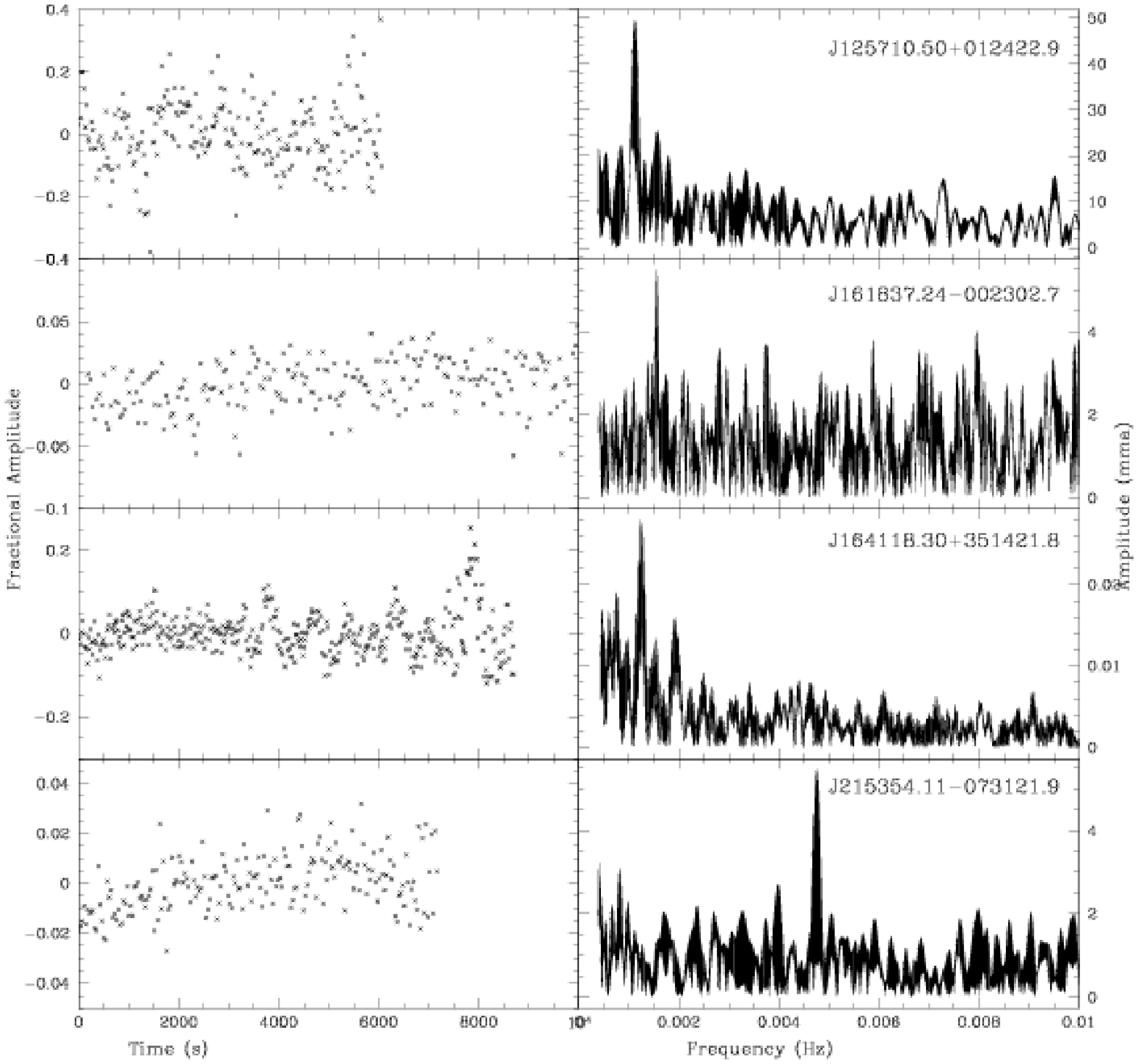}
\caption{The light curves of a single run (left panels) and the Fourier
transforms of the complete data set
(right panels) for the new ZZ Cetis, discovered at SOAR and McDonald
observatory.}
\label{lc2}
\end{figure}

In Fig.~\ref{lc} and \ref{lc2}, we show the light curves on the left panels 
and the Fourier transform (FT) on the right. We show only a small part of the
light curves, but the FTs were calculated for the entire dataset for each
target.
The lower detection limit in frequency is $2/T$, where $T$ is the total
length of the light curve, while the higher is the Nyquist frequency, 
$f_{\mathrm{Nyquist}}=2/t_{\mathrm{exp}}$.

The criterion we used to determine which peaks are real in the discrete FT
is to adopt an amplitude limit such that the probability of a noise peak 
exceeding this value is only 1/1000 (false alarm probability).
We then repeatedly subtracted the largest signal in the FT until there were
no more peaks above the detection limit (see Table~\ref{periodos}). 
The uncertainties in frequency are,
on average, 123$\mu$Hz, equivalent to 1s at P$\sim$100s.

\begin{table}
\begin{tabular} {|c|c|c|c|c|c|}
\hline
Star & $f$ ($\mu$Hz) & $P$ (s)  &  $A$ (mma) &  
$T_{\mathrm{max}}$ (s) & $T_0$ (BCT) \\
\hline \hline
HE 0031-5525 & 3035 & 329.5 & $2.5 \pm 1.0$ & $251 \pm 19$ & 2453231.6726750\\
& 3611 & 276.9 & $4.8 \pm 1.0$ & $245 \pm 13$ & \\
& 3638 & 274.9 & $1.5 \pm 1.0$ & $40 \pm 41$ & \\
\hline
SDSS J024922.3-010006.7 & 957 & 1045.2 & $10.9 \pm 1.5$ & $598 \pm 61$ & 
2453259.6947745\\
& 994 & 1005.6 & $5.6 \pm 1.5$ & $254 \pm 115$& \\
\hline
SDSS J125710.5+012411.9 & 1104 & 905.8 & $46.7 \pm 5.7$ & $31 \pm 18$ & 
2453526.525547\\
\hline
SDSS J153332.9-020600.0 & 3837 & 260.6 & $5.3 \pm 0.6$ & $130 \pm 14$
& 2453228.4215328\\
& 3879 & 257.8 & $4.3 \pm 0.6$ & $29 \pm 17$ &\\
\hline
SDSS J161837.2-002302.7 & 1553 & 644.0 & $5.4 \pm 1.1$ & $238 \pm 22$ & 
2453527.630436\\
\hline
SDSS J164115.5+352140.6 & 1276 & 809.3 & $27.3 \pm 4.6$ & $560 \pm 22$ & 
2453527.750116\\ 
\hline
SDSS J212808.4-000750.8 & 3309 & 302.2 & $17.1 \pm 1.3$ & $52 \pm 4$ & 
2453228.5198054\\
& 3638 & 274.9 & $11.0 \pm 1.3$ & $154 \pm 6$ &\\
& 3461 & 289.0 & $9.7 \pm 1.4$ & $131 \pm 7$ &\\
\hline
SDSS J213530.3-074330.7 & 1769 & 565.4 & $49.8 \pm 2.3$ & $12 \pm 5$ & 
2453228.6189298\\
& 1958 & 510.6 & $16.8 \pm 2.2$ & $18 \pm 14$ & \\
& 3094 & 323.2 & $13.0 \pm 2.3$ & $24 \pm 12$ & \\
& 3334 & 299.9 & $22.9 \pm 2.3$ & $209 \pm 6$ & \\
& 3549 & 281.8 & $13.3 \pm 2.3$ & $276 \pm 10$ & \\
\hline
SDSS J215354.1-073121.9 & 4757 & 210.2 & $5.6 \pm$ 0.9 & $15 \pm 5$ & 
2453527.816768\\
\hline
SDSS J223135.7+134652.8 & 1413 & 707.5 & $17.1 \pm 1.7$ & $389 \pm 16$ &
2453230.5708122\\
& 1595 & 627.0 & $26.3 \pm 1.7$ & $172 \pm 10$ &\\
& 1614 & 619.7 & $18.9 \pm 1.7$ & $104 \pm 13$ &\\
& 1822 & 548.7 & $13.7 \pm 1.7$ & $264 \pm 16$ &\\
& 2615 & 382.4 & $14.6 \pm 1.7$ & $369 \pm 10$ &\\
\hline
SDSS J230726.6-084700.2 & 825 & 1212.2 & $25.6 \pm 2.4$ & $478 \pm 32$ & 
2453228.6946388\\
& 1621 & 617.0 & $12.5 \pm 2.4$ & $457 \pm 33$ & \\
\hline
\end{tabular}
\caption{Periodicities identified in our data sets for the new ZZ Cetis, where
$f$ is the frequency, $P$ is the period, and $A$ is the amplitude. The times of 
maxima ($T_{\mathrm{max}}$) are given in relation to the $T_0$, in Barycentric 
Coordinate Time in days.} 
\label{periodos}
\end{table}

\begin{figure}
\centering
\includegraphics[angle=0,width=\linewidth]{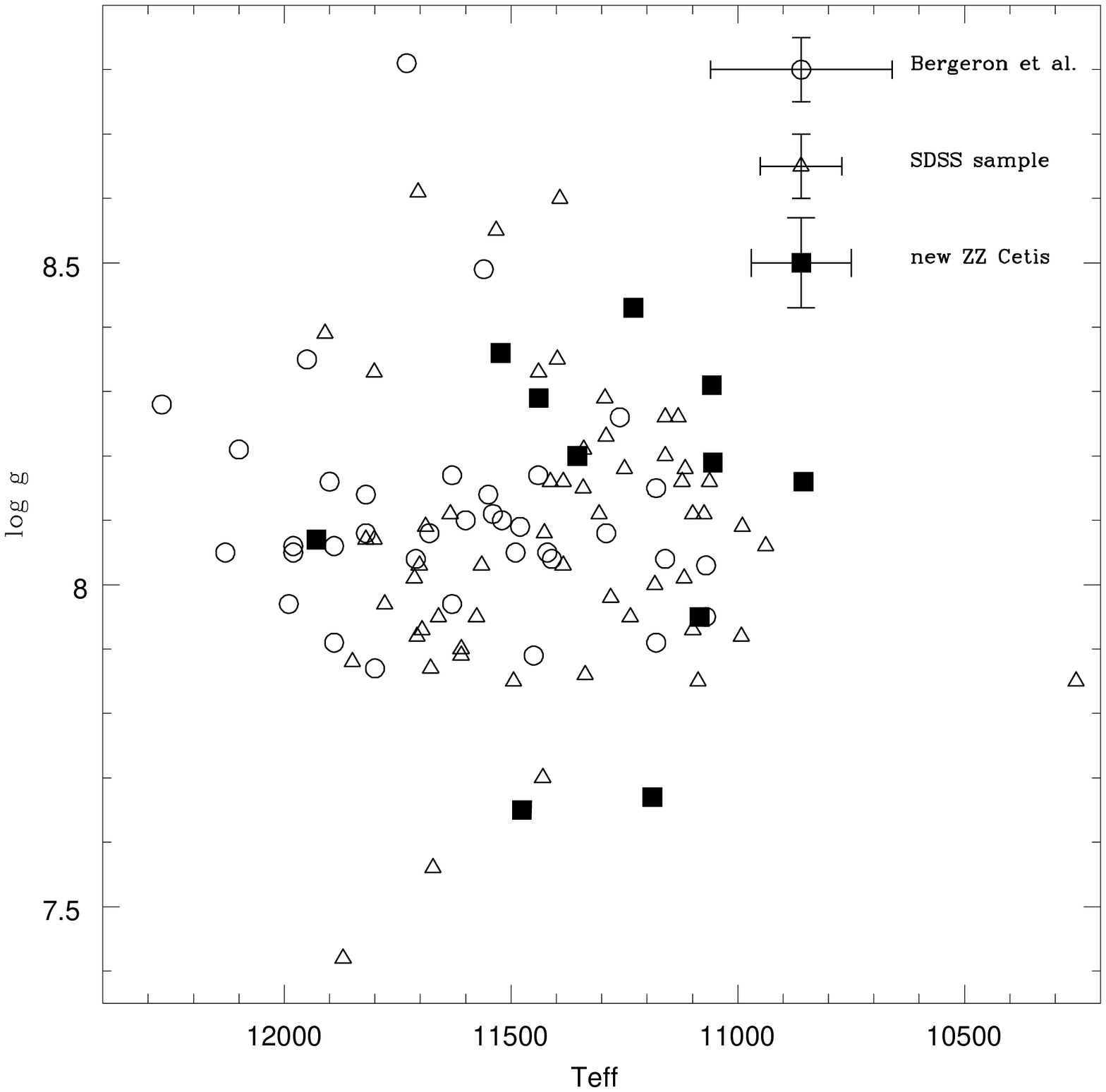}
\caption{The updated instability strip including all known ZZ Cetis. The 
circles are the stars described in Bergeron et al. (2004), the triangles are
the pulsators discovered by Mukadam et al. (2004), Mullally et al. (2005),
and Kepler et al. (2005b) from the SDSS, and the squares are the eleven new
ZZ Cetis presented in this paper. 
The average uncertainties for each set are on the top right of the plot.
}
\label{zzcetis}
\end{figure}

Combining these eleven new ZZ Cetis with the previously known pulsators, we see
a well defined instability strip in Fig~\ref{zzcetis}. The triangles represent
previously discovered candidates from SDSS, the circles are the stars fitted 
by Bergeron et al. (2004), and the squares are from this paper. It is
important to point out that this is not a homogeneous sample, as the 
previous determinations of $T_{\mathrm{eff}}$ and
$\log g$ were not obtained with the same model grid (e.g. Bergeron et al. 2004,
Koester \& Allard 2000), but the instability strip is restricted to a narrow
range of temperature 12\,270 K $\geq T_{\mathrm{eff}} \geq$ 10\,850 K. 
The pulsating star J235040.72-005430.9 (Mukadam et al. 2004) is clearly 
outside the 
ZZ Ceti 
instability strip. Despite this star being cooler than the
red edge of the instability
strip, its detected periods indicate it is a hot DAV. 
It is necessary to obtain a higher signal--to--noise spectrum, because
the SDSS spectrum has SNR$\sim$12, 
and analyze it with the line profile technique for an accurate
determination of both $T_{\mathrm{eff}}$ and $\log g$. 
Finally, it is necessary to increase the
sample of variables and non-variables at the borders of the instability
strip, to determine its exact boundaries. 

The star J214723.7-001358.4 did not show variability during 2 independent
runs, down to a limit of 3.75 mma. According to the SDSS spectrum, 
this star has $T_{\mathrm{eff}}=12000\pm280$ and 
$\log g=7.92\pm0.11$,
slightly hotter than the hottest ZZ Ceti from SDSS, reported in this
paper, J215354.1-074330.7 (see Table~\ref{newv} for details). 

However, the temperatures derived from SDSS optical spectra have external
uncertainties larger than 300\,K, as demonstrated from their duplicate spectra;
the uncertainties we quote in Table~\ref{newv} are the internal 
uncertainties in the
$\chi^2$ fitting, assuming no correlation between 
$T_{\mathrm{eff}}$ and $\log g$.
We conclude our observations are still 
consistent with a pure instability strip, but it does not exclude possible
contaminations, as the strip covers only $\sim$ 1200\,K. To solve this 
problem, we need spectra with SNR$\geq$50 to achieve 
$\sigma_{T_{\mathrm{eff}}} \leq 
200$~K and to
re-observe the stars which do not show variability in the literature to 
decrease their detection
limits below 4mma. Meanwhile, the purity of the ZZ Ceti instability strip
is still an open discussion.

Our results are mapping not only the blue edge, but also the red edge of the
instability strip. We discovered a pulsating star, J161837.2-002302.7, with 
$T_{\mathrm{eff}}$ and periodicities characteristics of the red edge, even
though the amplitudes are significantly lower
than expected. This star will be extremely important
as we try to understand how the ZZ Cetis will stop pulsating
around $T_{\mathrm{eff}}\sim10850$\,K.

\section{Concluding remarks}

We report the discovery of eleven new pulsating DA stars,
bringing the total to 106 known variables, all in the narrow
temperature range, 12\,270 K $\geq T_{\mathrm{eff}} \geq$ 10\,850 K,
corresponding to the
partial ionization of hydrogen and development of
a sub-surface convection zone. Three stars, HE~0031-5525,
SDSS J153332.9-020600.0 and SDSS J212808.4-000750.8, have pulsation 
characteristics
of hot DAVs, even though their derived $T_{\mathrm{eff}}$
are 11480 K, 11350 K, 11440 K, respectively. However, $T_{\mathrm{eff}}$ is not
the only parameter which determines the position of the instability strip. 
Giovannini et al. (1998) demonstrate that the instability strip also depends on
$\log g$. We therefore need to map the instability strip in both 
$T_{\mathrm{eff}}$ and $\log g$ using accurate determinations of these
parameters for the new variables.

\begin{acknowledgements}
Financial support: NASA origin grant, CAPES/UT grant, CNPq fellowship, DK and BV
acknowledge support from the Deutsche Forschungsgemeinschaft (DFG, grant
K0738/21-1).
\end{acknowledgements}

%
%

\end{document}